\documentclass[10pt,superscriptaddress,english,twocolumn,prl,showpacs,
				floatfix,aps]{revtex4-2} 
\usepackage[utf8]{inputenc}
\usepackage{amsmath}
\usepackage{amssymb}
\usepackage{graphicx}
\usepackage{xspace}
\usepackage{soul}
\usepackage{braket}
\usepackage[backref=none,bookmarksnumbered=true,bookmarks=true,
	bookmarksopen=true,colorlinks=true,citecolor=blue,linkcolor=blue,
	anchorcolor=green,urlcolor=blue,unicode=false]{hyperref}

\newcommand{\didv}{\ensuremath{\mathrm{d}I/\mathrm{d}V}\xspace}

\begin{document}

\title{Direct observation of intrinsic surface magnetic disorder in amorphous superconducting films}

\author{Idan Tamir}
\email{Idan.Tamir@FU-Berlin.de}
\affiliation{Fachbereich Physik, Freie Universit\"{a}t Berlin, 14195 Berlin, Germany.}
\author{Martina Trahms}
\affiliation{Fachbereich Physik, Freie Universit\"{a}t Berlin, 14195 Berlin, Germany.}
\author{Franzisca Gorniaczyk}
\affiliation{Department of Condensed Matter Physics, The Weizmann Institute of Science, Rehovot 76100, Israel.}
\author{Felix von Oppen}
\affiliation{Dahlem Center for Complex Quantum Systems and Fachbereich Physik, Freie Universit\"at Berlin, 14195 Berlin, Germany}
\author{Dan Shahar}
\affiliation{Department of Condensed Matter Physics, The Weizmann Institute of Science, Rehovot 76100, Israel.}
\author{Katharina J. Franke}
\affiliation{Fachbereich Physik, Freie Universit\"{a}t Berlin, 14195 Berlin, Germany.}

\date{\today}

\begin{abstract}
The interplay between disorder and interactions can dramatically influence the physical properties of thin-film superconductors. In the most extreme case, strong disorder is able to suppress superconductivity as an insulating phase emerges. Due to the known pair-breaking potential of magnetic disorder on superconductors, the research focus is on the influence of non-magnetic disorder. Here we provide direct evidence that magnetic disorder is also present at the surface of amorphous superconducting films. This magnetic disorder is present even in the absence of magnetic impurity atoms and is intimately related to the surface termination itself. While bulk superconductivity survives in sufficiently thick films, we suggest that magnetic disorder may crucially affect the superconductor-to-insulator transition in the thin-film limit.  
\end{abstract}

\maketitle
Disorder in superconducting thin films plays a decisive role in the determination of their physical properties. Although Anderson's theorem states that superconductivity is robust with respect to non-magnetic disorder \cite{anderson1959}, higher disorder, coupled with the ensuing enhancement of electron-electron interactions, does lead to an eventual reduction in the superfluid density and to local variations of the superconducting order parameter ($\Delta$) \cite{larkin1971,ma1985,sacepe2008}. Only at very strong disorder superconductivity breaks down and an insulating phase emerges at the superconductor-insulator transition (SIT) \cite{shahar1992,goldman1998,gantmakher2010,sacepe2020}.

In contrast, magnetic disorder has a much stronger impact on superconductivity. This comes about because magnetic scatterers break the time-reversal symmetry necessary for efficient superconducting pairing. As a result, even weak magnetic disorder is expected to lead to strong suppression of superconductivity \cite{abrikosov1960,tsang1980,chervenak1995}. The possible presence of 
large amounts of magnetic disorder is thus commonly disregarded when investigating disordered superconductors. Yet, a growing number of experimental observations indicate the presence of intrinsic magnetic surface disorder in various superconductors \cite{sendelbach2008,saveskul2019,yang2020}.

Direct evidence of magnetic surface disorder may be obtained by scanning tunneling microscopy (STM) and spectroscopy (STS) \cite{yazdani1997,heinrich2018}. Magnetic impurities exchange-coupled to a superconductor are expected to result in Yu-Shiba-Rusinov resonances inside the superconducting energy gap termed YSR states \cite{yu1965,shiba1968,rusinov1969}. So far, STS has been applied to a number of disordered superconductors \cite{sacepe2008,sacepe2011,chand2012,noat2013,szabo2016,lotnyk2017,liao2019,carbillet2020}. Most of these studies, using normal-metallic tips, reported on spatial variations in the width of the superconducting energy gap. Yet, YSR states have only been reported for amorphous Al and are ascribed to oxygen contamination \cite{yang2020}. 

Here, we use superconducting tips to spatially resolve variations in the spectra of the superconducting energy gap at the surface of amorphous indium oxide (a:InO) films, a frequently studied disordered superconductor \cite{hebard1990,shahar1992}. Our results are obtained from six different superconducting a:InO films with different degrees of disorder (for sample preparation and transport characterization see supplementary information, SI, \cite{SM}). The films have relatively high critical temperatures \cite{sacepe2015}, $T_C\gtrsim2.5$ K (well above our base temperature, $T=1-1.3$ K), as confirmed prior to the STM experiments by transport measurements (see SI for more details \cite{SM}).

\begin{figure}[ht!]
	\includegraphics  [width=8.5 cm] {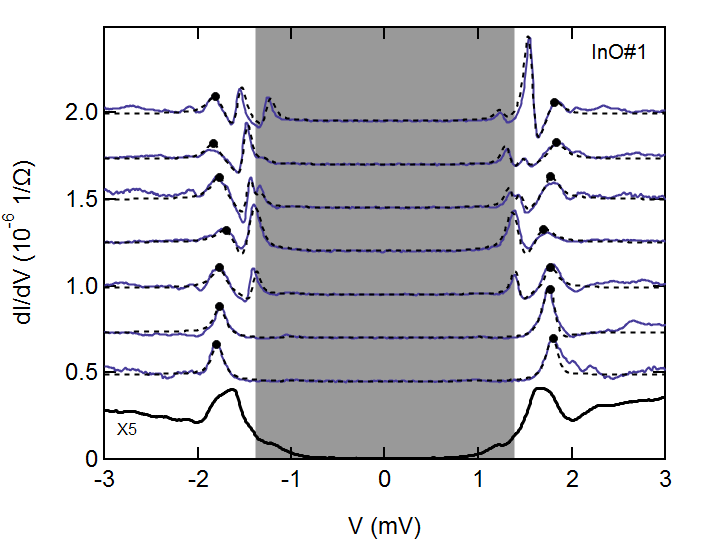}	
	\caption{\textbf{Observation of subgap states}. Representative local \didv vs. $V$ spectra (blue) observed by STS of an a:InO film. Traces are vertically shifted for visibility. The gray shaded region indicates $\Delta_{tip}$. Data are measured at $T=1.3$ K, and feedback opened at $I=200$ pA and $V=4$ mV, lock-in amplitude is 15 $\mu$V. The dashed black lines are fits to the data using a phenomenological model to include the effect of YSR states (for details see SI \cite{SM}). Black circles indicate the coherence-peak positions from the fits. The bottom-most trace (black, multiplied by five for visibility) is the average of all 400 evenly-spaced spectra measured within a $100\times 100$ nm\textsuperscript{2} scan range from which the representative spectra were selected. 
	}
	\label{ingap}
\end{figure}

We observe an abundance of subgap peaks on the surface of the a:InO films (see Fig. \ref{ingap}). In fact, if we also consider asymmetry in the height of the coherence peaks as an indication of weakly-bound subgap peaks \cite{homberg2020,odobesko2020}, we hardly observe any spectra that follow the classic predictions of the Bardeen-Cooper-Schrieffer (BCS) theory \cite{bardeen1957}. The response of these peaks to high-frequency radiation and tip approach suggests that these are YSR states \cite{farinacci2018,huang2020,chatzopoulos2021,peters2020}, and, although some experiments indicate the existence of surface magnetic disorder in similar systems \cite{hong2006,degraaf2017,mumford2021}, our results present, to the best of our knowledge, the first direct evidence. We also show that the presence of magnetic scattering centers is not due to surface contamination, in contrast to amorphous Al \cite{yang2020}, and therefore argue that it stems from the intrinsic presence of unsaturated bonds at the surface. Finally, while surface magnetic disorder should not suppress superconductivity in the bulk of the materials, we suggest that the role of surface magnetism will substantially increase when approaching the thin-film limit \cite{haviland1989}. The SIT may well be affected by the presence of surface magnetic disorder at this stage.

Our observations are made possible by utilizing superconducting tips (lead or niobium), shaped and characterized in ultra high vacuum to achieve atomic-scale spatial resolution. By using superconducting tips we overcome the thermal broadening limit ($\sim3.5~k_B T$ $\approx0.3$ meV at $T=1$ K, where $k_B$ is the Boltzmann constant), usually governing the energy resolution of spectroscopic measurements, and greatly improve the signal to noise ratio with respect to normal metallic tips. The measured spectra are then a convolution of sample and tip densities of states (DoSs) rather than being directly proportional to the sample's DoS. As a result, all of the sample related spectroscopic features are shifted in voltage ($V$) by the superconducting gap of the tip \cite{ruby2015} ($\Delta_{\text{tip}}/e=1.3-1.5$ mV, where $e$ is the electron charge, see SI for more details \cite{SM}). 

To survey the local superconducting properties, we record differential conductance (\didv) spectra on several macroscopically separated surface areas. Representative high-resolution spectra measured at different positions within one of these areas, in a range of $100\times 100$ nm\textsuperscript{2}, are presented in Figure \ref{ingap} (data collected on other samples is available in the SI \cite{SM}). Notably, most of the data do not resemble conventional BCS spectra. Instead, we often observe several peaks, symmetric in voltage, but asymmetric in intensity, and at varying subgap energies. The specific ratio of conventional BCS spectra to those exhibiting subgap peaks is sample and position dependent. For example, the data used in Fig. \ref{ingap} are selected from 400 evenly-spaced spectra, all measured within the same scan range, about $60\%$ of which exhibit well resolved subgap peaks.

We are able to capture the spectroscopic details of the data presented in Fig. \ref{ingap} by superimposing subgap quasi-particle states, at energies $\pm\varepsilon\le\Delta$, to the BCS form used to describe our sample's DoS. The fits to the spectra using this phenomenological model are plotted as black dashed lines in the figure (for full fit equation see SI \cite{SM}). Note that peaks inside the tip's gap (gray shaded region in Fig. \ref{ingap} are related to thermal occupation of quasi-particles \cite{ruby2015}, see SI \cite{SM}).

\begin{figure}[ht!]
	\includegraphics  [width=8.5 cm] {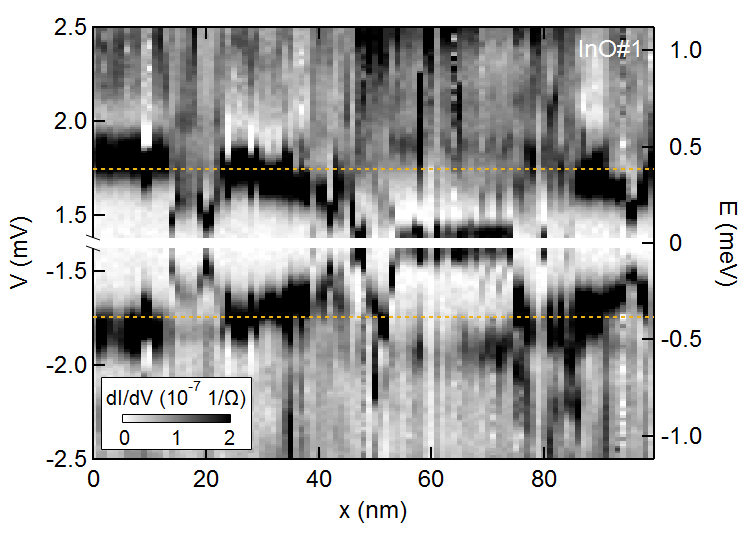}	
	\caption{\textbf{Spatial variations}. Color scale map of \didv spectra measured along a line. The orange dashed line indicates the averaged energy gap of the a:InO film. Right axis scale is corrected for $\Delta_{\text{tip}}=1.38$ meV. Data are measured at $T=1.3$ K, and feedback opened at $I=130$ pA and $V=2.5$ mV, lock-in amplitude is 15 $\mu$V. We note that the presence of a subgap excitation reduces the amplitude of the coherence peak. 
	}
	\label{linespecs}
\end{figure}

Next we probe the spatial extent of the subgap peaks. To do so we record densely (1 nm) spaced spectra along lines across the surface. An example is shown in Fig. \ref{linespecs}. We find a variety of subgap peaks at different energies, ranging from deep inside the a:InO superconducting energy gap to its edge, which is indicated by the dashed orange line (for clarity, we left out data at energies smaller than $\Delta_{\text{tip}}$, the complete spectra are available in the SI \cite{SM}). While some of the subgap peaks are rather local, others spread over a few nanometers. The spatial extent and abrupt changes in the spectra are clear evidence for the need for highly resolved data. We further emphasize the importance of high spatial resolution by including in Fig. \ref{ingap} the average of the entire set of spectra taken on the equidistant grid from which the representative data of that figure were selected (black, amplitude times five). The average's shape may easily be interpreted as a signature of BCS superconductivity with an unexplained, but not uncommon \cite{sacepe2008,chand2012,noat2013,szabo2016,sacepe2020}, non vanishing subgap conductivity.

\begin{figure*}[ht!]
	\includegraphics  [width=\linewidth] {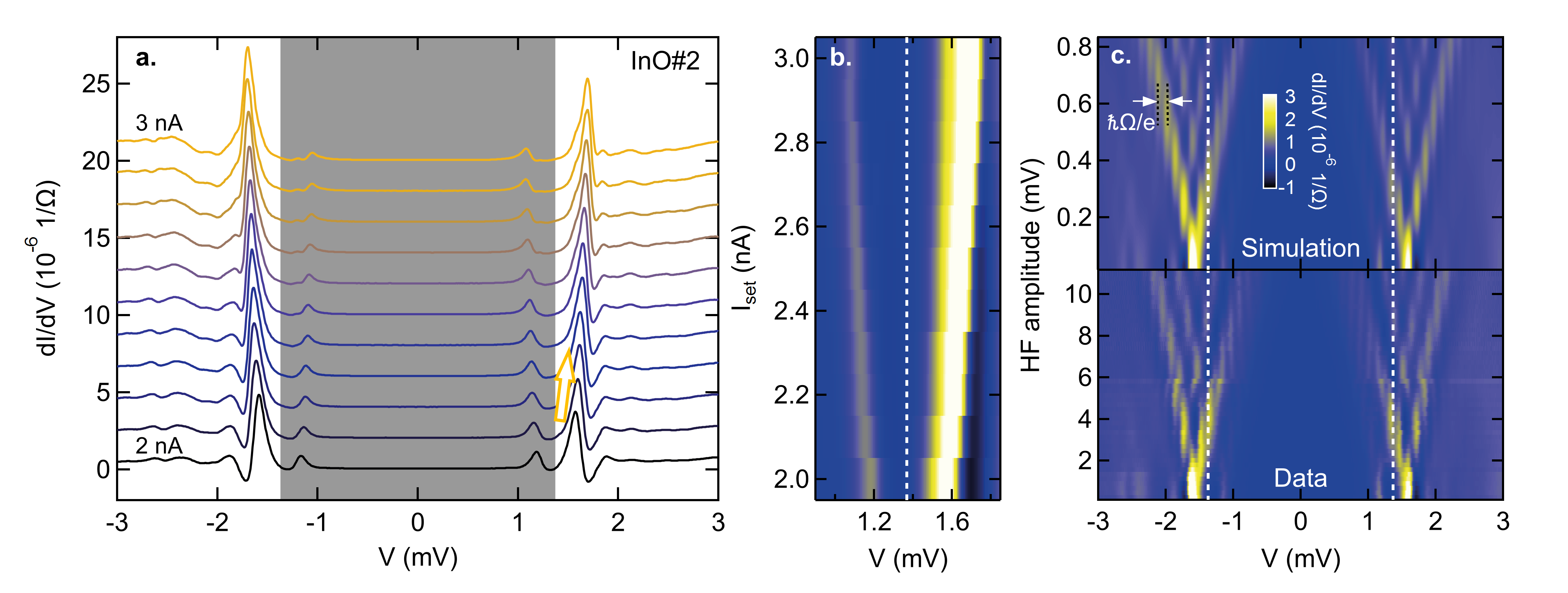}	
	\caption{\textbf{Approaching the surface and response to HF}. \textbf{a.} \didv vs. $V$ measured at different set currents from 2 nA (black) to 3 nA (orange) using 0.1 nA steps. The energy of the YSR state initially at $|\varepsilon+\Delta_{\text{tip}}|\sim1.5$ mV is shifted towards the coherence peaks that are initially barely resolved. The subgap peak's energy trend is indicated by the arrow. Spectra are vertically shifted for visibility. The gray shaded region indicates $\Delta_{tip}$. Data are measured at $T=1.3$ K, feedback is opened at $V=4$ mV, lock-in amplitude is 15 $\mu$V.  \textbf{b}. Color scale map of spectra using the same data as in \textbf{a}. For color scale, see \textbf{c}. Map focuses on positive $V$ for better visibility. Dashed line indicates $\Delta_{tip}$. The enhanced conductance below $\Delta_{tip}/e$ is related to the thermal replica of the $V<0$ YSR state. \textbf{c.} Color-scale map of \didv spectra measured at different HF power, bottom, and simulation assuming that single-electron tunneling dominates the transport ($\hbar\Omega/e$ splitting), top. Dashed lines indicate $\Delta_{tip}$. Data are measured at the same position, as in \textbf{a}, with the feedback opened at $V=4$ mV and $I=2$ nA. Differences in the HF amplitudes between measurement and simulation stem from damping in our transmission line.
	}
	\label{approach}
\end{figure*}

To gain further insight into the nature of the subgap states, we explore the influence of the STM tip on their excitation energy. Increasing the tunnel junction conductance by approaching with the STM tip toward the surface can result in an increase in the energy, $\varepsilon$, of the subgap states, as seen in Fig. \ref{approach}\textbf{a}., \textbf{b}. At even higher conductances, $\varepsilon$ saturates at the energy of the coherence peaks. Similar observations are made in other systems, where the subgap peaks are attributed to YSR states \cite{farinacci2018,huang2020,chatzopoulos2021}. Two possible mechanisms are suggested: First, pulling or pushing of a magnetic impurity due to van der Waals forces between tip and impurity, which in turn change the coupling between the magnetic impurity and the superconducting substrate \cite{farinacci2018,huang2020}. 
Second, local gating by the tip, which changes the electrostatic background \cite{chatzopoulos2021}. 
Due to the low tunnel-junction conductance used and the low charge density in our samples \cite{shahar1992}, we favor the latter explanation. Importantly, both explanations are compatible with a YSR origin of the observed subgap states. 

We note that the observed subgap states are not consistent with the recently reported collective gap in highly disordered superconducting a:InO films \cite{dubouchet2019}. In their report, Dubouchet \textit{et al.} observe subgap peaks that appear at high tunnel-junction transmission, larger than the conductance quantum ($G_0=2e^2/h$), and associate them with tunneling processes involving two electrons that indicate a collective energy gap of preformed Cooper pairs. Our samples, however, are far from the SIT, and our measurements are conducted in the low transmission limit. Furthermore, while only tunneling processes involving two electrons are relevant to the observation of a collective gap, we are able to determine the single particle nature of electrons tunneling into the subgap states we report here. This is done by irradiating the tunnel junction by high-frequency radiation (HF, $\nu$=$30$ GHz). Introducing radiation results in photon-assisted tunneling such that zero-radiation-power conductance peaks split in energy into several peaks with a separation inversely proportional to the excitation's charge $ke$, $\delta V=h\nu/ke$, where $h$ is Planck's constant and $k$ is an integer \cite{tien1963,roychowdhury2015,kot2020,peters2020,gonzalez2020}. 

A color-coded set of spectra with increasing HF amplitude is shown in Fig. \ref{approach}\textbf{c}, bottom panel. The pronounced YSR peaks are seen in white at the bottom of the map (lowest HF power). As we increase the HF power a splitting of the peaks is observed. Following Ref. \cite{peters2020} we simulated the data using only the spectrum measured without HF radiation and $\nu$ as input parameters. We find excellent agreement between experiment and the single-particle simulation ($k=1$).

The behavior of the subgap states in response to HF radiation and upon tip approach have similarly been observed for YSR states \cite{farinacci2018,huang2020,chatzopoulos2021,peters2020}, supporting this interpretation. However, Andreev bound states could potentially respond in a similar way \cite{sauls2018}. These might arise in regions where the superconducting gap is locally much reduced, confining electrons by Andreev reflections occurring at the boundaries. These bound states would also induce similar peaks in the measured \didv appearing symmetrically around the Fermi energy inside of the superconducting gap. However, our observation of states deep inside the superconducting energy gap would require a dramatic reduction of the superconducting gap in large parts of the surface. This is at odds with the previously reported gap variations of 10-20\% in moderately disordered films \cite{sacepe2008,szabo2016,carbillet2020}. 
Hence, we find this scenario unlikely and favor the interpretation in terms of YSR states. 

Having established the observation of spatially varying YSR states, we now discuss their magnetic origin. One may suspect that the YSR states are due to magnetic impurities arising from surface modifications induced, \textit{e.g.}, by surface oxide formation, or any other impurity absorption during storage. We took measures to remove several surface layers by sputtering. We set the sputtering parameters to remove approximately a 1 $\AA$ thick layer of surface atoms per minute, and conducted two sputtering cycles, 10 minutes each. For this experiment we used a thick sample (300 nm) to avoid thickness-dependent effects.

\begin{figure}[ht!]
	\includegraphics  [width=8.5 cm] {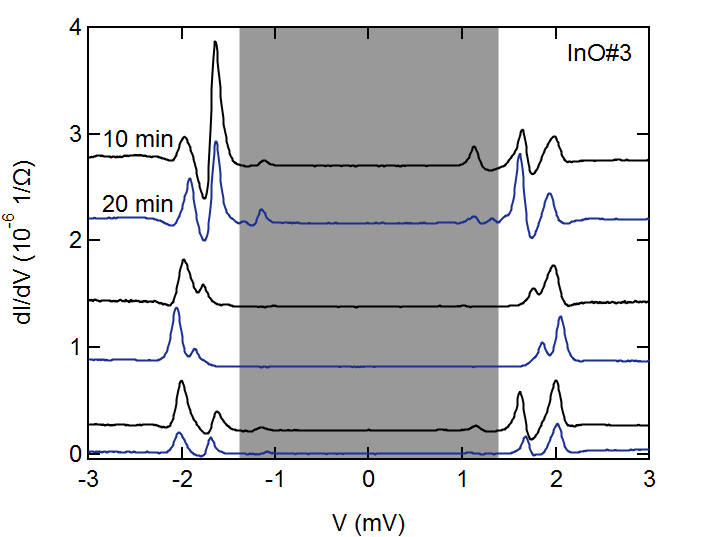}	
	\caption{\textbf{Removing surface layers}. Characteristic spectra, shifted for visibility, exhibiting pronounced YSR states. The spectra are measured after the first (black), and second sputtering cycle (blue). Data are measured at $T=1.3$ K, and feedback opened at $I=200$ pA and $V=4$ mV, lock-in amplitude is 15 $\mu$V. The gray shaded region indicates $\Delta_{tip}$.
	}
	\label{sputtering}
\end{figure}

Post sputtering we do not detect any qualitative change. In Fig. \ref{sputtering} we present spectra exhibiting similar YSR states after the first (black), and second sputtering cycle (blue). The spectra are shifted vertically for visibility. While we choose to present prominent subgap features, the entire sets of data acquired after sputtering are qualitatively very similar to each other and to data collected on other films without sputtering. Even after two sputtering cycles, the majority of individual spectra exhibit YSR states. We therefore conclude that the surface magnetic disorder is intrinsic and probably originates from unsaturated bonds at the surface. 

We emphasize that the YSR states are limited to the surface. A similar density of magnetic impurities in the bulk would lead to a strong suppression of superconductivity \cite{abrikosov1960,tsang1980,chervenak1995}. Nonetheless, we suspect that the observed intrinsic surface magnetic disorder can be of crucial importance for the thickness-driven SIT mainly due to enhanced exchange scattering as the sample thickness is reduced. %Scattering effects at the surface become increasingly relevant with reduced sample thickness, where exchange scattering may then affect the transition in addition to increased potential scattering. 
 
%\red{It is commonly believed that the oxygen content controls the degree of bulk disorder, and as such the SIT. Our results suggest that the number of unsaturated bonds, which can be tuned by the oxygen content at the surface, is of much stronger importance than previously anticipated. One may thus consider new strategies to steer the SIT with much simpler surface-preparation techniques, such as annealing in vacuum or oxygen atmosphere. However, our most important conclusion is that the previously unnoticed role of surface magnetic disorder should be considered in extended theoretical descriptions of the SIT.}

In conclusion, we showed that the previously unnoticed role of surface magnetic disorder should be considered in future experimental studies and theoretical descriptions of the SIT. Our results suggest that the number of unsaturated bonds, which can be tuned by the oxygen content at the surface, is of much stronger importance than previously anticipated. One may thus consider new strategies to steer the SIT with much simpler surface-preparation techniques, such as annealing in vacuum or oxygen atmosphere.

\section*{Acknowledgments}
We acknowledge financial support by Deutsche Forschungsgemeinschaft through grant CRC 183 (project C03) and JOSPEC (FR-2726/5). This research was supported by Grant No 2018024 from the United States-Israel Binational Science Foundation (BSF). I.T. acknowledges funding from the Alexander-von-Humboldt
foundation in the framework of Humboldt Research Fellowship
for Postdoctoral Researchers, and from the DFG in the framework of the Walter Benjamin Position.

\bibliographystyle{apsrev4-2}

\bibliography{ingap}

\end{document}